\documentclass{evn2002}
\begin{document}
   \title{Parsec Scale Properties of Radio Sources }

   \author{G. Giovannini\inst{1}
           \,
           M. Giroletti\inst{1}
           \,
           G.B. Taylor\inst{2}
           \,
           W.D. Cotton\inst{3}
           \, \\
           L. Lara\inst{5}
           \and
           T. Venturi\inst{4}
}

   \institute{Dipartimento di Astronomia, Universita' di Bologna, via Ranzani 
              1, 40127 Bologna, Italy
             \and
             NRAO, PO Box 0, Socorro NM 87801, USA
             \and
             NRAO, 520 Edgemont Road, Charlottesville, VA 22903-2475, USA
             \and
             Istituto di Radioastronomia, CNR, via Gobetti 101, 40129 Bologna,
               Italy
             \and
             Instituto de Astrofisica de Andalucia, CSIC, Apdo Correos 3004, 
               E-18080 Granada, Spain
             }

   \abstract{
We present a new complete sample of radio sources 
selected from the B2 and 3CR catalogues, with
no bias respect to the jet velocity and orientation. Using preliminary data
we investigate the parsec scale properties of radio sources with different
radio power and kpc scale morphology. We stress the evidence for high velocity
pc scale jets in all sources and conclude that the properties of parsec scale
jets are similar regardless of the source total power and large scale
morphology. Moreover we show two epoch images of two nearby radio galaxies:
a low power compact source and a CSO. The comparison of the two epoch images
suggests that in both sources the size is increasing with time. 
We derive a possible advance speed and estimate their dynamic age. 
   }

\authorrunning {Giovannini et al.}

   \maketitle
%

\section{Introduction}

The study of the parsec scale properties of radio galaxies is crucial to obtain
information on the nature of their central engine, and provides the basis of
the current {\it unified theories} (see e.g. Urry \& Padovani 1995), which
suggest that
the appearance of active
galactic nuclei strongly depends on orientation. 
 
To get new insight in the study of radio galaxies at pc resolution,
it is important to select source samples from low frequency catalogues,
where the source
properties are dominated by the unbeamed extended emission
and are not affected by observational biases related to orientation effects.
After the results presented in Giovannini et al. 2001, where we studied
a sample of 26
radio galaxies with strong cores,
we undertook a new project to observe a complete sample of radio
galaxies selected from the B2 and 3CR catalogs 
with z $<$ 0.1 and no limits
on the core flux density: the {\it Complete Bologna Sample} (CBS).
This sample consists of 95 sources and includes
23 of the 26 sources studied in Giovannini et al., 2001 (3C109, 3C 303, and
3C 346 are excluded because of the redshift limit). 
We observed with the VLBA 31 more sources allowing us to
discuss preliminary results using data from 54 on 95 sources.

Moreover we present observations at two different epochs (5 yrs apart)
of two peculiar sources of this sample
to study their possible expansion velocity and to derive their
dynamical age.
 
We use a Hubble constant H$_0$ = 50 km sec$^{-1}$ Mpc$^{-1}$ and
a deceleration parameter q$_0$ = 0.5

\section {Observations and source morphology}

We observed at 5 GHz
with the full VLBA and one VLA telescope\footnote
{VLBA and VLA are operated by NRAO as a facility of the NSF, operated 
under cooperative agreement by Assoc. Universities, Inc.}
31 sources from our sample to properly image
the parsec scale structure.
Each source was observed for about 1 hour with short scans at different 
hour angles to ensure a good uv-coverage. The observing data have been 
correlated in Socorro, NM. Postcorrelation processing used the NRAO AIPS 
package. All data were globally fringe fitted and then self-calibrated.
Final images were obtained using the AIPS and DIFMAP packages.

In Fig. 1 we show a plot of the observed core radio power versus total radio 
power for BCS sources. 
The line represents the correlation found by Giovannini et al. (1988), revised
according to Giovannini et al. (2001).
Sources show the well known dispersion around the straight line since the 
observed core radio power is affected by the source orientation angle.
Most sources with VLBI data are on the upper side with respect to the 
correlation line since 
observations are available only for $\sim$ half of the BCS sources and 
sources with a higher core
flux density have been observed so far.

   \begin{figure}
   \centering
   \vspace{170pt}
   \includegraphics{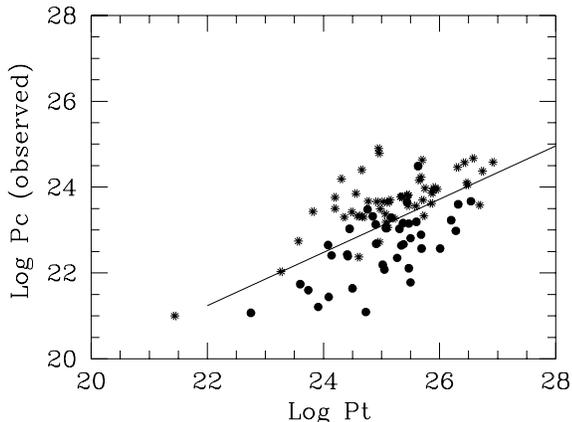}
      \caption{Total radio power at 408 MHz versus the measured
arcsecond core
radio power at 5 GHz for sources of the BCS.
Asterisks represent sources with VLBI data.
The straight line represents the correlation between core and total
radio power (Giovannini et al. 2001).
         }
   \end{figure}

Parsec scale structures
are mostly one-sided because of relativistic Doppler boosting effects.
However, some sources within the CBS have faint cores because of 
Doppler--deboosting. This implies that they are oriented at
sufficiently large angles to the line of sight so that counterjets are 
visible (see e.g. 3C 33 in Fig. 2).

   \begin{figure}
   \centering
   \vspace{230pt}
   \includegraphics{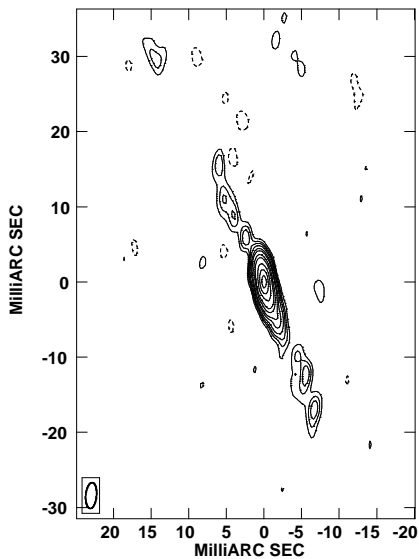}
      \caption{VLBA image of the narrow line FR II galaxy 3C33, at 5 GHz.
       Parsec scale jets are oriented as the extended kpc scale structure.
       Noise level is 0.1 mJy/beam. Levels are: -0.3 0.3 0.5 0.7 1 1.5 2 3 4 
6 8 10 mJy/beam.   }
   \end{figure}

\section{Parsec scale jet velocity}

There are several strong and widely accepted lines of evidences for the
existence of relativistic bulk velocities in the
parsec scale jets of radio sources: the observed super-luminal motions, the
rapid variabilities, the observed high brightness temperatures, the absence
of strong inverse-Compton emission in the X-ray and the
detection of a high frequency emission (gamma ray) for the
two BL-Lacs Mkn 421 and 501 all seem to require relativistic bulk speeds with
Lorentz factor $\gamma$ $>$ 3.

   \begin{figure}
   \centering
   \vspace{170pt}
   \includegraphics{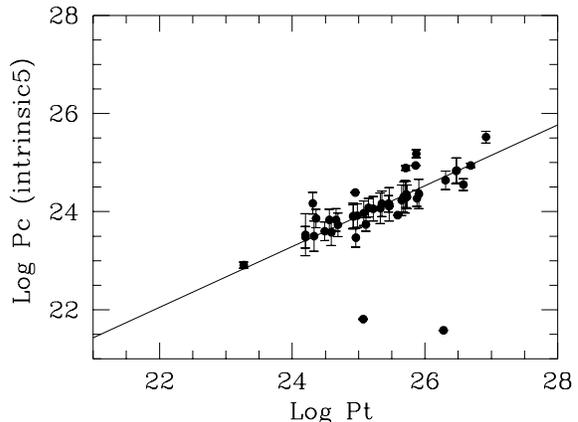}
      \caption{Total radio power at 408 MHz versus the intrinsic core
   radio power at 5 GHz derived with $\gamma$ = 5, for observed sources.
   The continuum line represents the correlation between the core and total
   radio power found by Giovannini et al., 2001. The two discrepant points
refer to the sources M87 and 3C192 (see text).
         }
   \end{figure}

We used the observational data to constrain the jet velocity and orientation. 
The methods used were the jet sidedness
and the core dominance (see Giovannini et al., 2001 for a more detailed
discussion). 

   \begin{figure*}
   \centering
   \vspace{330pt}
   \includegraphics{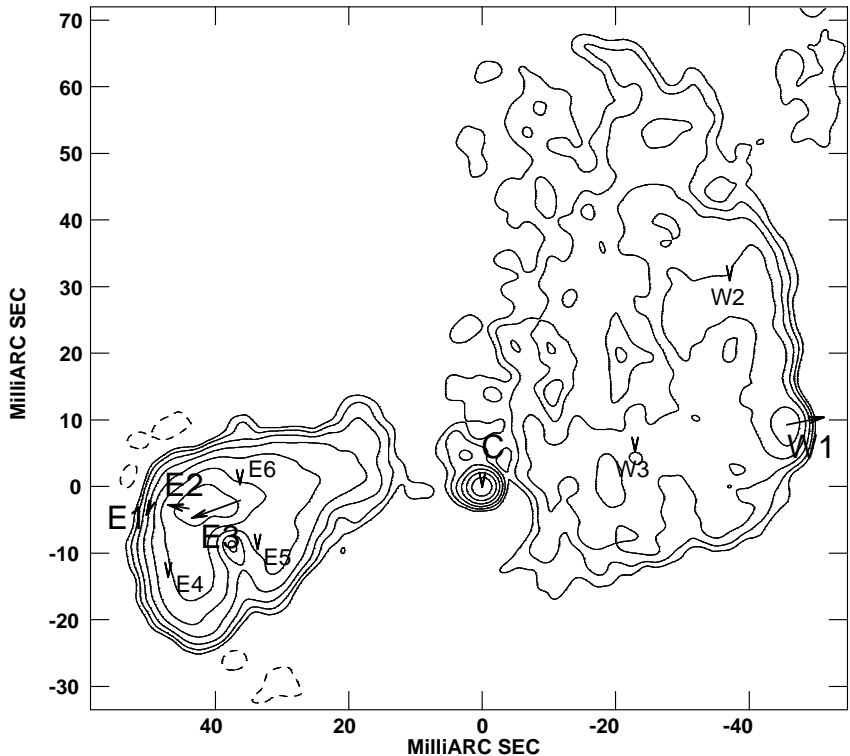}
   \caption{VLBI image of 4C31.04 at 5 GHz, July 2000 epoch. The HPBW is 3 mas,
            the noise level is 0.1 mJy/beam. Levels are: -0.4 0.4 0.8 1.6 3.2 
6.4 12.8 25.6 mJy/beam. The peak flux is 40 mJy/beam. Arrows indicate 
the positions
and motions of components derived from modelfitting, magnified by a factor 5.
At the source distance 1 mas = 1.56 pc          }
    \end{figure*}

The derived constraints confirm that
in all sources radio jets move at high velocities on the mas scale.
Since in many cases
we can only give a lower limit to $\gamma$, we used the following approach:
we assumed different
$\gamma$ values and we tested if the derived source properties
are in agreement with the observational data.
Once a jet velocity is assumed, the jet orientation is constrained by the
observational data and it is possible to compute the corresponding Doppler
factor $\delta$ = ($\gamma (1 - \beta cos \theta))^{-1}$ for each
source.
Then,
from the value of $\delta$ and of the measured
radio power, we can derive the intrinsic core radio power for each source:
P$_{c-observed}$ = P$_{c-intrinsic}$ $\times$ $\delta^2$ (assuming $\alpha$
= 0).
Since there is a range of possible jet orientations, we have a possible
range of values for $\delta$ and therefore of P$_{c-intrinsic}$.

We found that $\gamma$ cannot be larger than 10, or we would see a 
dispersion in the
plot of the observed core versus total radio power 
larger than reported in Giovannini et al., 1988 (see also
Fig. 1).

Moreover observational data rule out values of $\gamma$ lower 
than 3. Such low values imply too small Doppler factor
corrections, and we still see the effect of different orientation angles
in the distribution of core radio power. This is in agreement with the 
evidence of high velocity jets discussed at the beginning of this Section.

Assuming 3 $<$ $\gamma$ $<$ 10 and plotting the intrinsic core radio power 
versus
the total radio power, we find that all sources but two (M87 and 3C192), are
in good agreement with the correlation line (see Fig. 3)
and have a small dispersion around it,
as expected since the spread due to the different orientation 
angles is removed.
This result implies similar jet velocities ($\gamma$ = 5 in Fig. 3) for all
sources despite the variety of their large scale morphology and different 
total radio power. We 
remind that a correlation between the core and total radio power is 
expected if sources are in energy equipartition conditions,
as discussed in Giovannini et al. (2001).

The radio core of the two discrepant sources (M87 and 3C 192) is too faint
with respect to their total
radio power at 408 MHz. 
In order to shift M87 on the correlation
we need parsec scale jets moving at a much higher velocity than 
the other sources and/or an orientation angle larger than the
angle derived from its high proper motion (see Giovannini et al. 2001). 
For 3C 192, even assuming that this source is on the plane of the sky, we
need very high velocity parsec scale jets, 
alternatively we are observing a core in a low activity
phase: if it is turning off, this source could be in a pre-relic phase.

Present data from 54 BCS sources confirm the results obtained by Giovannini
et al. (2001) on the basis of a subsample of 26 radio sources. In particular we
confirm that the observational data imply similar jet velocities, 
with $\gamma$ in the range 3 -- 10 for all
sources, in spite of the variety of their large scale morphology and different 
total radio power. 

\section{4C31.04}

The radio galaxy 4C31.04 (0116+31) was tentatively classified as a low redshift
(z = 0.0592) 
Compact Symmetric Object (CSO) by Cotton et al., 1995. Conway (1996) showed
the presence of a complex HI absorption across the lobes. 
A detailed optical study is given by Perlman et al. (2001).
Giovannini et al. (2001) confirmed the CSO structure and the core 
identification with the faint flat
spectrum component in between the two extended lobes. Here we present a second 
epoch observation of this source with the VLBA and one VLA telescope 
obtained
on July 03, 2000, five years after the first epoch image at the same frequency.

   \begin{figure}
   \centering
   \vspace{200pt}
   \includegraphics{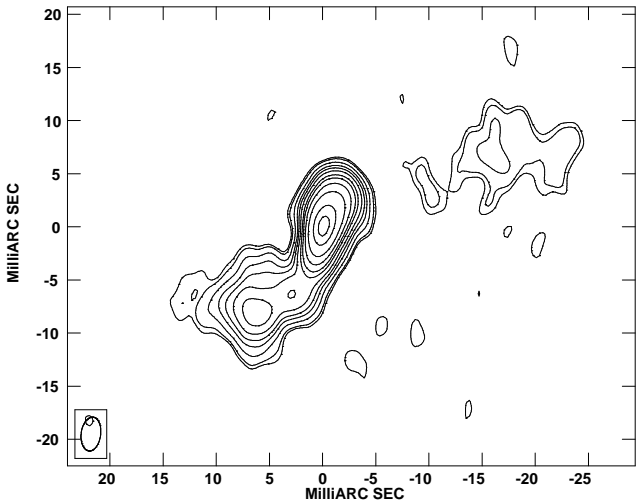}
      \caption{VLBI image at 5 GHz of NGC 4278, July 2000 epoch. The noise 
level
is 0.04 mJy/beam. Levels are: 0.1 0.15 0.3 0.5 0.7 1 1.5 2 3 5 10 15 mJy/beam. 
The HPBW is 3.2 $\times$ 1.8 mas in PA -7$^\circ$.
 At the source distance 1 mas = 0.06 pc         }
   \end{figure}

The images at the two epochs were calibrated and reduced in the same way 
and were compared 
to look for proper motion. The two images are in very
good agreement, and the same radio features are visible in both of them.
We measured the position of several structures
visible in the images with respect to the core using different methods,
e.g.
by model-fitting (DIFMAP), by fitting gaussian components
with JMFIT (in AIPS) and by subtracting the second epoch image from the first
one. 
In fig. 4 we show the final image at 5 GHz, July 2000 epoch, with superposed
arrows, whose length is proportional to the apparent motion, derived from 
the comparison
of the two epochs. Despite the low brightness complex structure of the two
lobes, their expansion is well defined in the peaks 
and confirmed by the shift of the whole sharp edges (well 
visible in the {\it difference} image).

The measured shift is $\sim$ 0.54 $\pm$ 0.1 mas on the West side; 
on the East side a region moved of $\sim$ 0.7 $\pm$ 0.2 
mas while in a different region we measured an expansion of $\sim$ 0.3 
$\pm$ 0.07 mas.

We are aware of the uncertainty present in our analysis,
made using only two epoch data, 
however we note that the relative position of the core 
is well defined and that both lobes are increasing their
distance from it. The source expansion
is evident as a general increase of the distance of the
whole lobes, as clearly visible in the subtraction image (Giroletti et al.,
in preparation), and it is confirmed by the measured shift of a few 
substructures.
Of course new observations to be obtained in a few years
are necessary to confirm this result.

If we take into account that the source is near the plane of the sky 
(Giovannini et al., 2001), the measured shift implies  
an average advance speed
of $\sim$ 0.5c in both lobes. Assuming a constant 
velocity we derive a source kinematic age of only $\sim$ 400 yrs.
This result strongly supports the model where the radio emission in CSO objects
arises from a recently activated radio source. 
Even considering possible
temporal variations in the lobe advance speed, the age estimate (which should 
be considered a lower limit) is expected to be within a factor 10 of the 
real source age (Owsianik and Conway, 1998).
The lobe advance speeds are higher but still in agreement with previous
velocities measured in CSO sources (Perlman et al., 2001 and references in).

\section{NGC 4278}

We observed this nearby (z = 0.0021)
elliptical galaxy at two different epochs spaced by 5 years
with the VLBA + Y1 at 5 GHz (as 4C31.04). This radio galaxy is a low power
radio source (Log P$_{tot}$ = 21.44 W/Hz at 408 MHz). The radio morphology
visible in our VLBI images (see Fig. 5) is quite different from 4C31.04, as
expected from the different radio power. In NGC 4278 a central flat spectrum
core is visible with two elongated lobes in S-E (the main one) 
and N-W direction. These features are not well collimated but
reseambles extended jets in low power FR I sources.
We note that all the source flux density measured in arcsecond scale VLA
observations at the same frequency, is present in our VLBI image excluding
the existence of a larger scale emission.
 
We compared the two epoch images using the same technique as in 4C31.04.
Also in this case the {\it difference} image between the two epochs and
the comparison of substructure positions with model-fitting 
show that the blob at the end of 
the SE component moved between the two epochs of $\sim$ 0.84 $\pm$ 0.2 mas 
corresponding at the
source distance to an advance speed $\sim$ 0.03c $\pm$ 0.01c.
Assuming a constant expansion velocity, the dynamic age of this source
is very low, of the order of only 100 yrs. We consider this value  
only as indicative, given that only two epochs were
used in the comparison. 
A low expansion velocity is expected from the low radio power and the
source morphology. In this case we do not have a hot spot expanding in
the Interstellar Medium, but a low power jet slowly increasing its size.
We expect that a low power radio source as NGC 4278 will not grow to a giant
radio source, but that it will become a small size low power radio source,
as e.g. NGC 5322 (Feretti et al. 1984). We suggest that NGC 4278 is 
a young low power radio source, slowly growing to become a low power 
small size radio galaxy.
 
\begin{acknowledgements}

The European VLBI Network is a joint facility of European, Chinese and
other
radio astronomy institutes funded by their national research councils.
This research was supported by the European Commission's IHP Programme
``Access to Large-scale Facilities", under contract No.\ HPRI-CT-1999-00045
We acknowledge the support of the European Comission - Access to Research
Infrastructure action of the Improving Human Potential Programme.
We acknowledge the support of the European Union Infrastructure
Cooperation Network in Radio Astronomy, RadioNET. 

We thank Dr. L. Feretti for useful discussions and comments.

\end{acknowledgements}


\begin{thebibliography}{}

\bibitem[1996]{} Conway, J.E. 1996, in IAU Symp. 175, R.D. Ekers, C. Fanti,
                 \& L. Padrielli (Dordrecht; Kluwer), p92

\bibitem[1995]{} Cotton, W.D., Feretti, L., Giovannini, G., Venturi, Lara, L.
                 1995, ApJ, 452, 605

\bibitem[1984]{} Feretti, L., Giovannini, G., Hummel, E., Kotanyi, C.G. 1984
               A\&A, 137, 362
 
\bibitem[1988]{} Giovannini, G., Feretti, L., Gregorini, L., Parma, P. 1988,
           A\&A, 199, 73


\bibitem[2001]{} Giovannini, G., Cotton, W.D., Feretti, L., Lara, L., Venturi, 
                T. 2001, \apj, 552, 508

\bibitem[1998]{} Owsianik, L., Conway, J.E. 1998, A\&A, 337, 69

\bibitem[2001]{} Perlman, E.S., Stocke, J.T., Conway, J., Reynolds, C. 2001,
                \aj 122, 536

\bibitem[1995]{} Urry, C.M., \& Padovani, P. 1995 \pasp, 107, 803

\end{thebibliography}
\end{document}